\documentclass[fleqn,usenatbib]{mnras}
\usepackage{newtxtext,newtxmath}
\usepackage{changepage}
\usepackage[T1]{fontenc}

\DeclareRobustCommand{\VAN}[3]{#2}
\let\VANthebibliography\thebibliography
\def\thebibliography{\DeclareRobustCommand{\VAN}[3]{##3}\VANthebibliography}

\title[Betelgeuse with MATISSE]{Images of Betelgeuse with VLTI/MATISSE across the Great Dimming}

\usepackage[usenames]{color}
\usepackage[dvipsnames]{xcolor}

\usepackage{enumitem}
\usepackage{graphicx}
\usepackage{hyperref}
\usepackage{float}
\usepackage{tabularray}
\usepackage{footnote}
\usepackage{comment}
\usepackage{tabularx}
\usepackage[utf8]{inputenc}
\usepackage[para,flushleft]{threeparttable}

\definecolor{green}{rgb}{0.,1.,0.}
\definecolor{blue}{rgb}{0.,0,1}
\definecolor{orange}{rgb}{1,0.5,0}
\definecolor{red}{rgb}{1,0.,0}

\newcommand\FM[1]{{\color{Green} #1}}  % FlorentinFlorentin
\newcolumntype{Y}{>{\centering\arraybackslash}X}

\author[J. Drevon et al.]{
J. Drevon$^{1,2}$,
F. Millour$^{1}$,
P. Cruzal\`ebes$^{1}$,
C. Paladini$^{2}$,
P. Scicluna$^{2}$,
A. Matter$^{1}$,
A. Chiavassa$^{1}$,
\newauthor
M. Montarg\`es$^{3}$, %%% Ask Miguel the list of co-authors
E. Cannon$^{4}$,
F. Allouche$^{1}$,
K.-H. Hofmann$^{5}$, 
S. Lagarde$^{1}$,
B. Lopez$^{1}$,
A. Meilland$^{1}$,
\newauthor
R. Petrov$^{1}$,
S. Robbe-Dubois$^{1}$,
D. Schertl$^{5}$,
G. Zins$^{2}$,
P.~Ábrahám$^{6}$,
P.~Berio$^{1}$,
Th.~Henning$^{7}$,
\newauthor
J. Hron$^{8}$,
J. W. Isbell$^{7}$,
W. Jaffe$^{9}$,
L.~Labadie$^{10}$,
J.~Varga$^{9,6}$,
G. Weigelt$^{5}$,
J.~Woillez$^{11}$,
R.~van~Boekel$^{7}$,
\newauthor
E.~Pantin$^{3}$
W.C. Danchi$^{12}$,
A. de Koter$^{13}$,
V. G\'amez Rosas$^{9}$,
M.R.~Hogerheijde$^{9,14}$,
J.~Leftley$^{1}$,
\newauthor
P.~Stee$^{1}$,
R.~Waters$^{15,16}$
\\ 
\\
$^{1}$Universit\'e Côte d'Azur, Observatoire de la C\^ote d'Azur, CNRS, Laboratoire Lagrange UMR 7293, B\^atiment H. Fizeau, F-06108 Nice Cedex 2, France \\
$^{2}$European Southern Observatory, Alonso de Córdova, 3107 Vitacura, Santiago, Chile \\
$^{3}$LESIA, Observatoire de Paris, Universit\'e PSL, CNRS, Sorbonne Universit\'e, Universit\'e Paris Cit\'e, 5 place Jules Janssen, F-92195
Meudon, France\\
$^{4}$Intitute of Astronomy, KU Leuven, Celestijnenlaan 200D B2401, 3001 Leuven, Belgium\\
$^{5}$Max-Planck-Institut f\"ur Radioastronomie, Auf dem H\"ugel 69, D-53121 Bonn, Germany \\
$^{6}$Konkoly Observatory, Research Centre for Astronomy and Earth Sciences, H-1121 Budapest, Hungary \\
$^{7}$Max Planck Institute for Astronomy, K\"onigstuhl 17, D-69117 Heidelberg, Germany \\
$^{8}$Department of Astrophysics, University of Vienna, T\"urkenschanzstrasse 17, Austria\ \\
$^{9}$ Leiden Observatory, Leiden University, Niels Bohrweg 2, NL-2333 CA Leiden, The Netherlands \\
$^{10}$I. Physikalisches Institut, Universit\"at zu K\"oln, Z\"ulpicher Str. 77, D-50937 K\"oln, Germany \\
$^{11}$European Southern Observatory, Karl-Schwarzschild-Str. 2, D-85748 Garching, Germany \\
$^{12}$NASA Goddard Space Flight Center, Astrophysics Division, Greenbelt, MD 20771, USA \\
$^{13}$Anton Pannekoek Institute for Astronomy, University of Amsterdam, 1090 GE Amsterdam, The Netherlands \\
$^{14}$Anton Pannekoek Institute for Astronomy, University of Amsterdam, Science Park 904, NL-1090 GE Amsterdam, The Netherlands \\
$^{15}$Department of Astrophysics/IMAPP, Radboud University, P.O. Box 9010, NL-6500 GL Nijmegen, The Netherlands \\
$^{16}$SRON Netherlands Institute for Space Research Sorbonnelaan 2, NL-3584 CA Utrecht, The Netherlands \\
}
   
\date{Accepted XXX. Received YYY; in original form ZZZ}

\pubyear{2015}

\begin{document}

\label{firstpage}
\pagerange{\pageref{firstpage}\pageref{lastpage}}
\maketitle
\begin{abstract}
    From Nov. 2019 to May 2020, the red supergiant star Betelgeuse experienced an unprecedented drop of brightness in the visible domain called the great dimming event. Large atmospheric dust clouds and large photospheric convective features are suspected to be responsible for it. To better understand the dimming event, we used mid-infrared long-baseline spectro-interferometric measurements of Betelgeuse taken with the VLTI/MATISSE instrument before (Dec. 2018), during (Feb. 2020), and after (Dec. 2020) the GDE. We present data in the 3.98 to 4.15\,$\mu$m range to cover SiO spectral features molecules as well as adjacent continuum. We have employed geometrical models, image reconstruction, as well as radiative transfer models to monitor the spatial distribution of SiO over the stellar surface.  We find a strongly in-homogeneous spatial distribution of SiO that appears to be looking very different between our observing epochs, indicative of a vigorous activity in the stellar atmosphere. The contrast of our images is small in the pseudo-continuum for all epochs, implying that our MATISSE observations support both cold spot and dust cloud model.
\end{abstract}

  % % context heading (optional)
  % % {} leave it empty if necessary  
  %  {CONTEXT}
  % % aims heading (mandatory)
  %  {AIMS}
  % % methods heading (mandatory)
  %  {METHOD}
  % % results heading (mandatory)
  %  {RESULTS} 
  %  %locate where the dust is present, and where the C$_2$H$_2$+HCN molecules are lying.}
  % % conclusions heading (optional), leave it empty if necessary 
  %  {CONCLUSION}

% Select between one and six entries from the list of approved keywords.
% Don't make up new ones.
\begin{keywords}
(stars:) massive - (stars:) imaging - (stars:) individual - (stars:) variables: general - techniques: interferometric - techniques: image processing 

\end{keywords}

\section{Introduction}

\textbf{The red supergiant star Betelgeuse --}\ 
Betelgeuse ($\alpha$~Orionis) is an O-rich star \citep{Perrin2007} with a sun-like metallicity \citep{Meynet2013} and a mass-loss rate of 1.2$\times$10$^{-6}$\,M$_\odot$yr$^{-1}$ \citep{LeBertre2012}. This red supergiant (RSG) of spectral type classified as M2I by \citet{Levesque2005} has a mass estimated between 16.5 and 19\,M$_\odot$ by \citet{Joyce2020}. The same authors give a distance of $168^{+28}_{-15}$\,pc using seismic analysis, while \citet{Harper2017} give $222^{+48}_{-34}$\,pc using new radio data taken with e-MERLIN, and older ALMA and VLA data combined with the Hipparcos catalogue \citep{vanLeeuwen2007}. Its effective temperature T$_{\rm{eff}}$ is estimated to range from 3650$\pm$25\,K, using spectro-photometric observations \citep{Levesque2005}, to 3690$\pm$54\,K \citep{Ohnaka2014}, using VLTI/AMBER spectro-interferometric observations. It is also a semi-regular variable star (SRc) with two main periods of $\approx$\,400 days and $\approx$\,2000 days identified by \citet{Gray2008}. \\ 

\textbf{The great dimming scenario --}\ 
From Nov. 2019 up to May. 2020, Betelgeuse went through an unexpected V-magnitude increase up to an historical value of $1.614\pm0.008$ \citep{Guinan2020} reached between the 7th and the 13th of Feb. 2020. In this paper, we  refer this 6 months event as the "great dimming event" (GDE). The proposed steps of the scenario for the GDE are: \\

\textbf{Step I: Formation of a large plasma outflow [Feb. 2018--Jan. 2019] --} Deep successive shocks generate a huge convective outflow gradually bringing the material to the surface. \citet{Kravchenko2021} detect two shocks located beneath the stellar photosphere, the stronger one in Feb. 2018 and the weaker one in Jan. 2019, using high resolution spectroscopic observations (R=86000) between 0.38 and 0.90\,$\mu$m with HERMES \citep{Raskin2011} at La Palma. The second shock, increasing the effect of the first one, contributes to the progressive formation of a plasma outflow at the surface of the photosphere; \\

\textbf{Step II: Hot spot formation [Sep.--Nov. 2019] --} The material growing up at the surface of the photosphere creates a hot bubble. \citet{dupree_2020} show an increase of the UV signal in the lines and in the continuum from Sept. to Nov. 2019 using Hubble observations. They interpret it as the sign of a hot, dense, luminous and high-temperature structure located in the southern hemisphere of the star between the photosphere and the chromosphere; \\

\textbf{Step III: Cold spot and dust formation [Nov. 2019--Feb. 2020] --} The new material detached from the photosphere forms progressively a gas cloud above the surface. Because of the missing material, a colder area appears on the photosphere seen as a dark spot in the line of sight and it allows the formation of dust in part of the cloud located above it. Spectro-photometric observations between 0.4 and 0.68\,$\mu$m, made by \citet{Levesque2020} to estimate the effective temperature gradient induced by the GDE. They show no temperature variation between March 2004 and Feb. 2020, while \citet{Harper2020b} show a decrease of 125\,K from Sept. 2019 to Feb. 2020 between 0.719 and 1.024\,$\mu$m. They explain these two apparently inconsistent results because of the presence of an opaque area made of cooler material covering at least more than half of the stellar surface. In addition, the presence of a dark spot at the surface of the star coupled with extinction of dust located above the surface agree with the VLTI/SPHERE visible observations made by \citet{Montarges2021} and VLTI/MATISSE N-band observations made by \citet{Cannon2023}. It is also in good agreement with observations in polarized visible light of \citet{Cotton2020}. They show a decrease of the polarization degree between Oct. 2019 and Feb. 2020, and explain it as a decrease of asymmetric illumination of the stellar surface and/or changes in the dust distribution. 

\begin{table}
\begin{minipage}[]{0.5\textwidth}
\caption{Observation logbook.}
\resizebox{.94\textwidth}{!}{%
\begin{tabularx}{\linewidth}{c c c c c c c}
\hline
 Date & Timespan  & Cals. & AT-config. & Seeing &  $\tau_0$  & OBs.$^{a}$ \\
  &   &  &  &  ["] &  [ms] &  \\
\hline
\multicolumn{7}{l}{2018-12-}  \\
03 & 04:35--07:55 & 1$^{b}$, 2$^{c}$& A0B2J2C1 & 0.9 & 2.9 & 12\\
05 & 03:49--04:14 & 2 & A0B2J2C1 & 0.7 & 1.5 & 6  \\
08 & 01:45--07:29 & 1, 2 & A0B2D0C1 & 0.6 & 3.5 & 24  \\
09 & 03:30--07:53 & 1, 2 & K0B2D0J3 & 0.4 & 4.8 & 17  \\
11 & 03:27--07:52 & 1, 2 & K0G2D0J3 & 0.6 & 8.7 & 24  \\
12 & 04:30--06:16 & 1 & A0G1D0J3 & 0.6 & 5.7 & 8   \\
14 & 04:04--07:23 & 1 & A0G1J2J3 & 0.6 & 5.6 & 8  \\
15 & 04:02--05:50 & 1, 2 & A0G1J2K0 & 0.5 & 5.8 & 12  \\
\hline
\multicolumn{7}{l}{2020-02-} \\
08 & 00:01--02:01 & 1 & A0B2D0C1 & 0.6 & 5.9 & 12  \\
19 & 00:19--01:37 & 2 & K0B2D0J3 & 0.9 & 3.5 & 6  \\
\hline
\multicolumn{7}{l}{2020-12-} \\
14 & 04:10--07:21 & 1, 2 & A0G1J2J3 & 0.5 & 10.1 & 24  \\
18 & 04:01--05:17 & 1 & K0G2D0J3 & 0.6 & 5.6 & 12  \\
27 & 01:56--03:31 & 2 &  A0B2D0C1 & 0.8 & 5.0 & 18  \\
\hline
\end{tabularx}%
}
\footnotesize
\begin{tablenotes}
    \item $^a$ Number of observing blocks actually used for the reduction and calibration \\ of the data, $^b \varepsilon$~Lep (K4III, $\theta_L$=5.7$\pm$0.5\,mas), $^c \alpha$~CMa (A1V, $\theta_L$=5.8$\pm$0.5\,mas).
\end{tablenotes}
\label{tab:log}
\end{minipage}
\end{table}

For our study, we use L-band data taken with the VLTI/MATISSE instrument \citep{Lopez2022} in medium spectral resolution (R$\approx$500) before, during, and after the GDE to probe and monitor the geometry of the close circumstellar environment of Betelgeuse. The present letter is organized as follows: i) in Section~\ref{sec:Observation} we present new VLTI/MATISSE observations of Betelgeuse in L-band; ii) in Section~\ref{sec:Spec_analysis} we introduce the spectral analysis and the uniform disk fitting; iii) in Section~\ref{sec:Images} we show and compare the first reconstructed images of the SiO environment of Betelgeuse for the three epochs; and iv) in Section~\ref{sec:Discussion} we present our interpretations and conclusions.

\begin{comment}
\begin{itemize}
    \item in Section~\ref{sec:Observation} we present new VLTI/MATISSE observations of Betelgeuse in L-band, and describe the data, the processing and the calibration steps;
    \item in Section~\ref{sec:Spec_analysis} we introduce the spectral analysis and the uniform disk fitting;
    \item in Section~\ref{sec:Images} we show and compare the first reconstructed images of the SiO environment of Betelgeuse for the three epochs;
    \item in Section~\ref{sec:Discussion} we present our conclusions.
\end{itemize} 
\end{comment}

\section{Observation and data reduction} \label{sec:Observation}

The three observing epochs of Dec. 2018, Feb. 2020, and Dec. 2020 are located before, during, and after the GDE of Betelgeuse ( The N band data of the GDE already shown in \citet{Cannon2023}). All the MATISSE observations are close to a light curve photometric minimum of variability.

We reduced and calibrated the observation data using the MATISSE data reduction pipeline and  additional python tools provided by the MATISSE consortium\footnote{mat\textunderscore tools v.0.1 freely available on Gitlab \url{https://gitlab.oca.eu/MATISSE/tools/-/tree/master}}. The pipeline provides calibrated oifits files of the science target comprising measurements of the spectral distributions of flux, squared visibility, differential and closure phases, in the LM- and N-bands of the mid-infrared. 
Finally, we reject each night contaminated by adverse instrumental or weather effects, not already identified by the MATISSE pipeline. In Table~\ref{tab:log} we list only the remaining nights after our sorting.

\begin{comment}
    To increase the accuracy of the interferometric calibration, we re-estimate the angular diameter of the calibrator targets in such a way that the spectral distribution of the visibility transfer function becomes as flat as possible over the considered spectral range. Using this flatness criterium on the estimated transfer function in the L-band thanks to the MATISSE observed visibilities we find uniform disk diameter values of 5.7$\pm$0.5\,mas for $\varepsilon$~Lep, and 5.8$\pm$0.5\,mas for $\alpha$~CMa, while the JSDC v2 catalog of \citet{Bourges2014} gives 5.9$\pm$0.6\,mas and 6.1$\pm$0.5\,mas, respectively.\\
\end{comment}

\begin{comment}
\section{Description of the data} \label{sec:data_desc}

 \FM{Furthermore, the deviation from a constant as a function of wavelength can be used } to re-estimate the diameter of resolved calibrators (7 Cet, Sirius, and eps Lep) and refine the transfer function curve for the calibration part of the data reduction.   

Figure~\ref{fig:FIT} top panel shows the Betelgeuse spectrum in the $3.98 - 4.15 \mu m$ range with the SiO first overtone molecular bands superimposed over a pseudo-continuum. These lines are weak (c.a. 20\% of the pseudo-continuum), and in absorption. The third panel shows the MATISSE squared visibilities (indicators of the object's size) as a function of wavelength fometer baseline (projected baseline $B_{p}\approx 9 m$). 
\end{comment}

\section{Spectral Analysis and Uniform Disk fitting} \label{sec:Spec_analysis}

\begin{figure}
    \centering
    \includegraphics[width=0.48\textwidth]{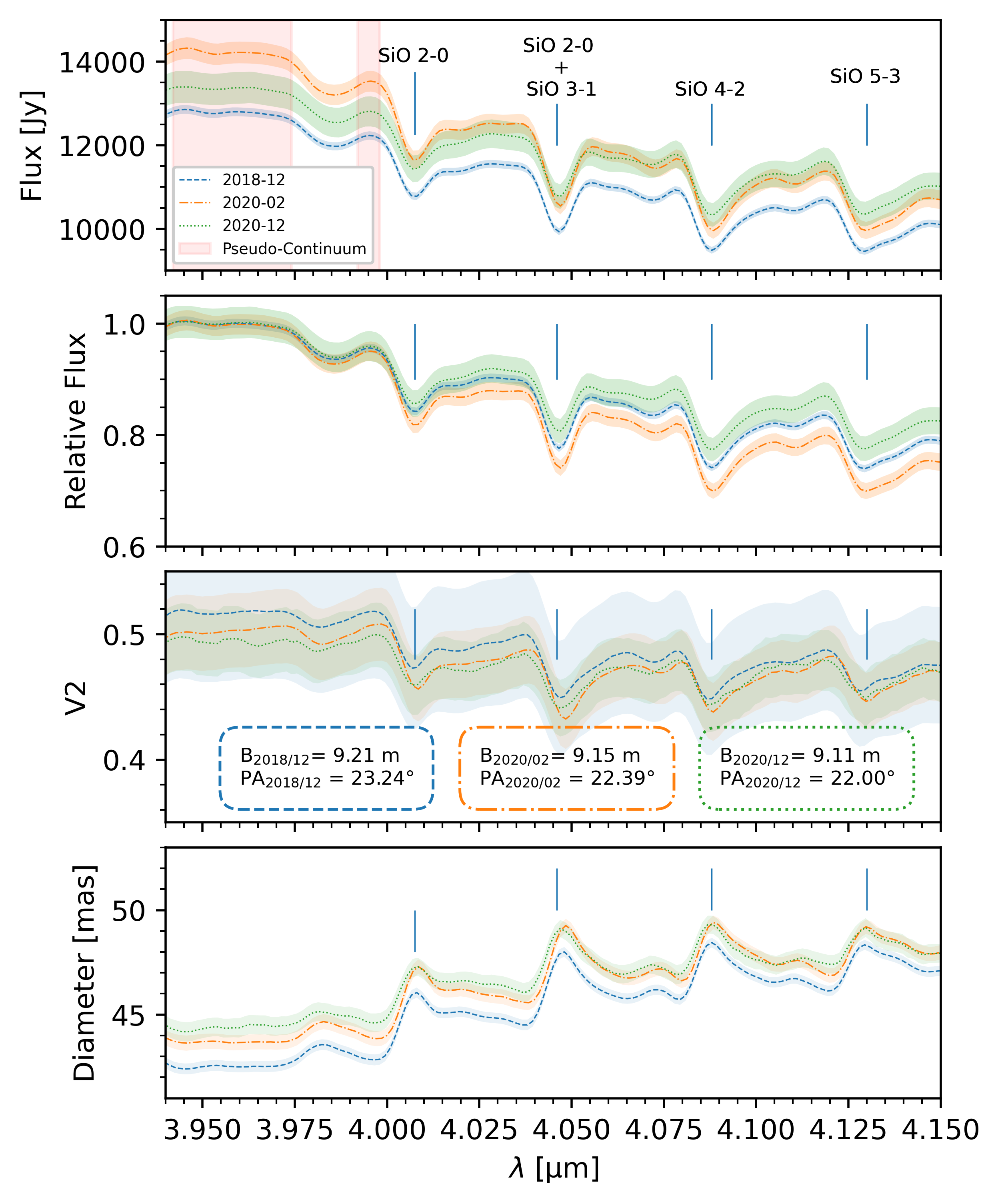}
    \caption{Top panel: VLTI/MATISSE absolute spectra for the three epochs with the identification of the main features.  The filled area close to the data point corresponds to the error bars associated to the given quantities. The red area represent the pseudo-continuum range used in this work. Second panel: Relative flux with respect to the continuum for the three epochs. Third panel: visibility squared plotted versus wavelengths for the various epochs of the observations. Bottom panel: fitted uniform disk diameter versus wavelengths for the various epochs.}
    \label{fig:FIT}
\end{figure}

Figure~\ref{fig:FIT} top panel shows a comparison between the three epochs of the Betelgeuse spectrum in the 3.98--4.15 $\mu$m range. The SiO lines are weak, $\approx$~20\% of the pseudo-continuum , and are in absorption. We observe changes in the spectrum between the 3 epochs. The pseudo-continuum level defined in this work between 3.942--3.974~$\mu$m and 3.992--3.998~$\mu$m seems to be brighter than pre-dimming by $\approx$~10\% during the GDE, and by $\approx$~5\% post-dimming. In the second panel of the figure we remark that the normalized flux to the continuum show SiO absorption features located at 4.01$\mu$m, 4.05$\mu$m, 4.09$\mu$m, and 4.13$\mu$m \citep{Aringer1997} which are slightly deeper during the Great Dimming (Feb. 2020) than before and after the event (Dec. 2018 and 2020).
In the third panel we show the squared visibility for similar baselines and position angles for the three epochs. We observe small differences between the various epochs, these are within the measurement uncertainties, therefore it is hard to conclude a trend.
In the bottom panel of Figure~\ref{fig:FIT}, we show the fitted equivalent uniform disk diameter obtained. The diameter's variation for a given epoch reflects the presence of one or several layers of material above the stellar photosphere. The equivalent visibility-fit diameter appears smaller in Dec. 2018, while the star appears larger during and after the dimming event. This change has not to be interpreted as an actual change of the star's photospheric diameter, but rather a change of brightness of the atmosphere layers (see also Fig.~\ref{fig:IMAGES}). In addition, the  error bars associated to the angular diameter look smaller regarding the errors of the squared visibilities shown above, due to the fact that the first-lobe uniform disk fitting is constrained by  dozens of squared visibilities measurements (and not only the values shown in the third panel of Figure~\ref{fig:FIT}).

\section{Images} \label{sec:Images}

\begin{figure*}
    \centering
    \includegraphics[width=\textwidth]{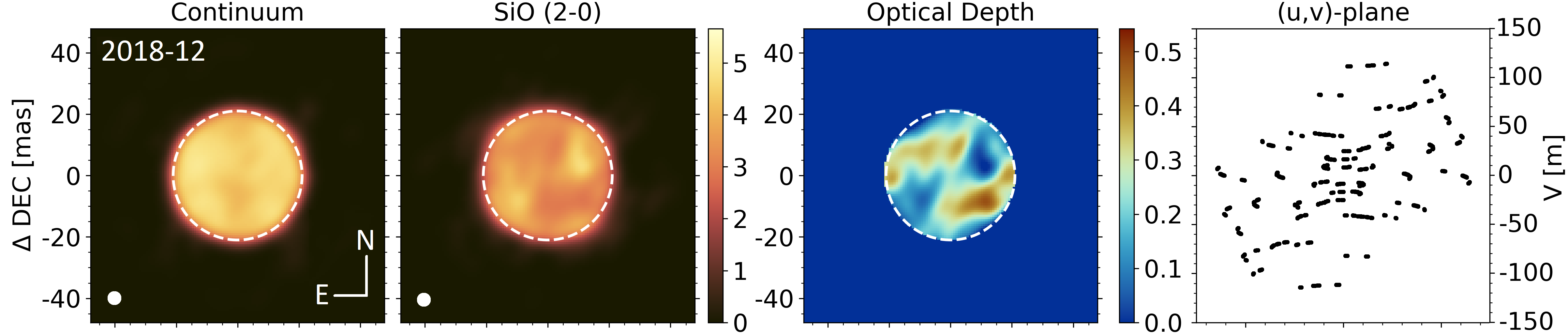}
    \includegraphics[width=\textwidth]{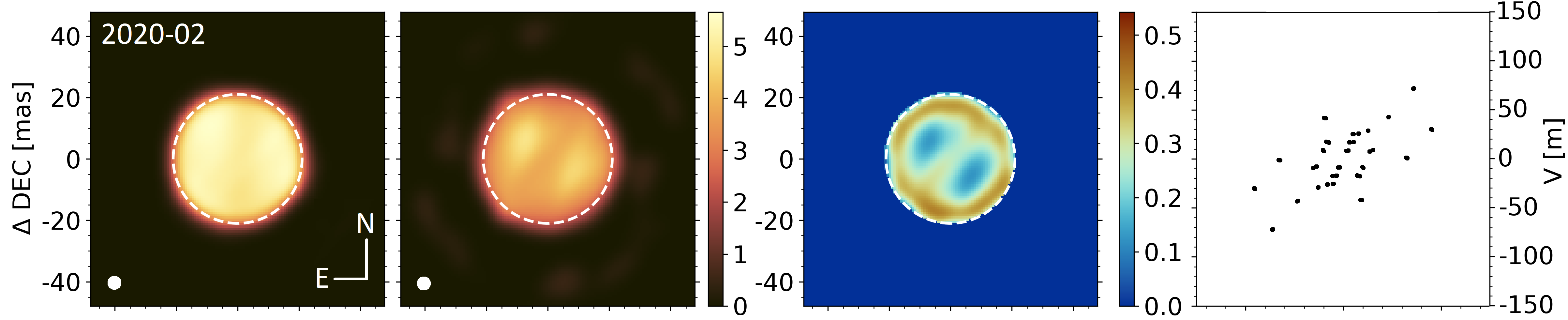}
    \includegraphics[width=\textwidth]{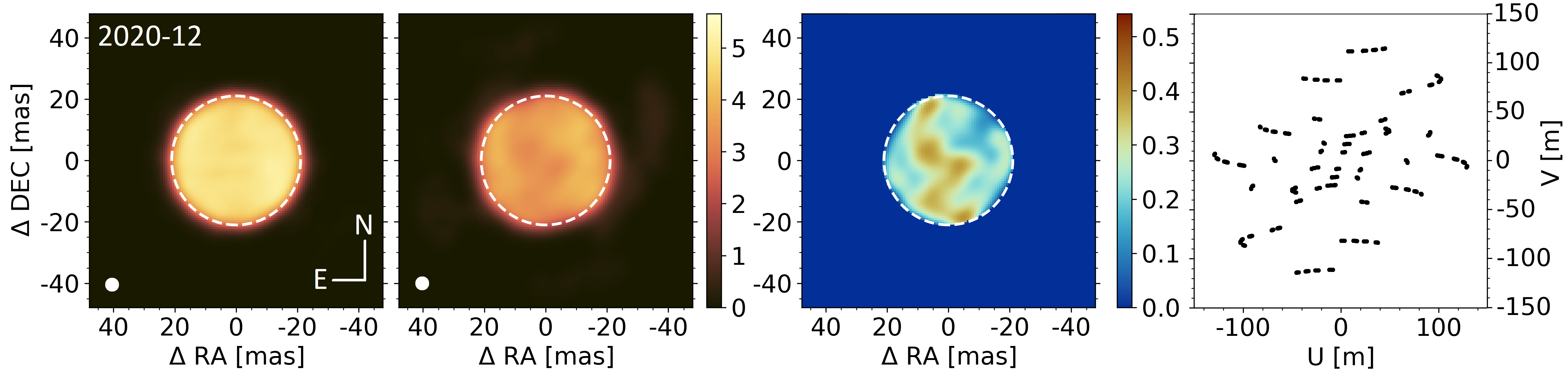}

    \caption{Each rows correspond to a given epoch precised in the top-left corner of the first panel. From left to right: 1) Reconstructed map in the pseudo-continuum, 2) in the SiO (2--0) absorption band (both scaled in Jy/pix with a squared pixel size of 0.78\,mas), 3) deduced map of the optical depth for the SiO (2--0), 4) (u,v)-plane coverage of the used measurements. The dashed white circles on the maps enclose intensities higher than 70\% of the maximum (42\,mas in diameter). The small white disks at the bottom left corners of the reconstructed maps corresponds to an equivalent circular interferometric beam used for the convolution of all the reconstructed images ($\approx$~4 mas in diameter). The angular size has been estimated using the Feb. 2020 observations which has the poorest (u,v)-plane coverage among the three epochs.}\label{fig:IMAGES}
\end{figure*}

We use the image reconstruction software IRBIS \citep{Hofmann2014} to produce L-band maps at the three epochs of observation in the combined spectral ranges 3.942--3.974~$\mu$m and 3.992--3.998~$\mu$m (pseudo-continuum), and in the spectral range 4.004--4.012~$\mu$m (SiO 2--0 transition). After a cautious scan of the imaging parameter space, we found the following optimal parameters for the reconstruction: i) initial image: an uniform disk, ii) prior: maximum entropy, and  iii) hyper-parameter: $\mu$ = 10. We show the obtained images in Figure~\ref{fig:IMAGES}. Since the Feb. 2020 observation run has a sparse (u,v)-plane coverage, in order to compare the three epochs images together, we convolved them with the Feb. 2020 equivalent circular interferometric beam of an angular diameter of 4\,mas.

In order to make a quantitative analysis between the reconstructed maps obtained at the 3 epochs, we compute the values of the photometric median, the median absolute deviation (MAD), and the intensity contrast as defined by \citet{Wedemeyer2009}. The values reported in  Table~\ref{tab:image_values} are computed for intensities higher than 70\% of the maximum \citep{Chiavassa2022} to exclude the region of the disk dominated by the limb darkening effect. The corresponding size measures about 42\,mas, shown as dashed white circles in the reconstructed maps of Figure~\ref{fig:IMAGES}. In addition, we use the equations developed in Appendix~\ref{sec:depth_map} to estimate the optical depth map of SiO, shown in the center-right panels of Figure~\ref{fig:IMAGES}. This allows us to monitor the SiO gas distribution across the three epochs. 

\begin{table}
\begin{minipage}[t]{0.48\textwidth}
\caption{Photometric quantities of the reconstructed maps. $\delta$I is the intensity contrast as defined by \citet{Wedemeyer2009} estimated with an error bar of 6\%, and MAD is the median absolute deviation of the intensity. The squared pixel size is 0.78\,mas.}
\centering
%\resizebox{1.1\textwidth}{!}{%
%\setlength\tabcolsep{1.5pt} % default value: 6pt
\begin{tabularx}{0.95\textwidth}{ @{}cYccc@{}}
\hline
 Date & Band & $\delta$I & Median & MAD \\ 
  & & & [Jy/pix] & [Jy/pix] \\ 
 \hline
2018-12 & Pseudo-cont.  & 0.12 & 4.4 & 0.3 \\ %& 0.4 \\
2020-02 & Pseudo-cont.  & 0.14 & 5.1 & 0.5 \\ %& 0.6 \\
2020-12 & Pseudo-cont.  & 0.12 & 4.7 & 0.3 \\ %& 0.4 \\
2018-12 & SiO (2--0) & 0.14 & 3.4 & 0.3 \\ %& 0.5 \\
2020-02 & SiO (2--0) & 0.16 & 3.8 & 0.4 \\ %& 0.5 \\
2020-12 & SiO (2--0) & 0.11 & 3.6 & 0.3 \\ %& 0.4 \\
\hline
\end{tabularx}%
%}
\label{tab:image_values}
\end{minipage}
\end{table}

\section{Discussion} \label{sec:Discussion}

We comment here the 4\,$\mu$m maps reconstructed from the VLTI/MATISSE instrument obtained before, during, and after the GDE in the SiO (2--0) absorption band and in the continuum.

\subsection{ Wavelength and Time variations of the angular diameter}

%Assuming that the increase of the fitted uniform disk diameter corresponds to a change of the photospheric stellar diameter and not a change in the brightness distribution of the stellar atmosphere, then 

The uniform disk angular diameter, obtained in section \ref{sec:Spec_analysis}, exhibit chromatic variations. These variations give hints to the geometry of the SiO dust seed formation region of Betelgeuse. Indeed, for example in Dec. 2018, while the apparent uniform disk diameter is $\approx$42.5\,mas in the pseudo-continuum, it is $\approx$46.0\,mas for the SiO 2-0 transition and and $\approx$48\,mas for the remaining 3 transitions (SiO 2-0 + SiO 3-1, SiO 4-2, and SiO 5-3), meaning that, qualitatively, the SiO layer extends 2\,mas above the stellar photosphere, and spans 2\,mas thick. Of course, such a qualitative analysis should be supplemented by more quantitative ones, using state-of-the-art molecular models of Betelgeuse atmosphere to constrain the vertical structure of the SiO distribution, but it is out of the scope of this letter.

The UD angular diameter shows also time variations: figure~\ref{fig:FIT} indicates a global increase of 3\% of the visibility-fitted uniform disk angular diameter between Dec. 2018 and Feb. 2020. From 42.5\,mas to 43.8\,mas in the continuum and 46.0\,mas to 47.4\,mas in the SiO (2--0) transition band, this small increase is about the same order of magnitude as the 1\% increase reported in K-band \citep{Montarges2021}. However, as explained by \citet{Dharmawardena2020} and \citet{Montarges2021}, such a small variation of the angular diameter is not sufficient to explain the GDE.\\

\subsection{Time variation of the pseudo-continuum}
\label{sec:varPseudoCont}

The intensity contrast gives important quantitative information about the distribution of the surface brightness intensity of the star. Its uncertainty value of 6\% has been estimated by measuring the variation of the intensity contrast on reconstructed images from simulated interferometric data of a uniform disk (see images in figure~\ref{fig:IRBIS_test}, and more details in section~\ref{sec:simIRBIS}). A temporal intensity contrast comparison between different images allows us to probe the surface and atmospheric activities between the different periods. In Table~\ref{tab:image_values} we report an increase of the intensity contrast over the surface in the pseudo-continuum images around 4\,$\mu$m of $\sim$~17\% between Dec. 2018 (before the GDE) and Feb. 2020 (during the GDE). Coupling those quantities with a slight increase of the median flux, this transcribe the appearance of a global brighter surface during the GDE which might be associated to recently dust formed in the line of sight. Figure~\ref{fig:FIT} also shows a global increase of the infrared emission in the continuum of $\sim$\,13\% which is in agreement with the presence of dusty material recently formed close to the star as proposed by \citet{Montarges2021}. The results of \citet{Harper2020a}, who do not find any infrared excess around 25\,$\mu$m, are not necessarily in disagreement with our results and might be complementary to constraint the nature of the chemical composition of the newly formed dust in the environment of the star during the GDE (but this last point is out of the scope of the current work).

%This is in agreement with \citet{Cotton2020} who underline a similar decrease in the polarization observations. 
%We also observe a dark region extending from South-East to North-West, also present in the continuum, which can be associated to a cold and dark region at the surface potentially leading to the formation of dust in the line of sight. 

\subsection{Time variation of the SiO asymmetries}

Since the quality and the robustness of the image reconstruction strongly depend on the completeness of the (u,v)-plane coverage, we need to be careful about our interpretation on the structures shown in the images, especially when the (u,v)-plane is poor like the Feb. 2020 one as shown in Figure~\ref{fig:IMAGES}. Indeed, the presence of artifacts in the images may impair our interpretation. They are flux whose owns a poorly constraint shape due to the holes in the (u,v)-plane coverage. 

Assuming that dust seeds precursors such as SiO molecule can be used as tracers for dust formation region, therefore asymmetries in the gas distribution can also be a way to probe asymmetries in the distribution of recently formed dust. In Dec. 2018 and Dec. 2020, we observe strong and different asymmetries in the SiO (2--0) gas distribution. This is in agreement with some episodic and non-homogeneous ejection of mass expected for a RSG \citep{Toala2011}. Regarding the Dec. 2020 images, we cannot exclude artifacts contamination induced by poor coverage of the (u,v)-plane. Therefore we cannot make any strong conclusions about changes in the gas distribution during this epoch.\\

%\subsection{Variation of the SiO (2--0) band}

In the SiO (2--0) band, the intensity contrast increases during the GDE by 14\% with an increase of the median flux of 10\% with respect to Dec. 2018. Therefore, it seems that during the GDE we observe brighter structures in the line of sight. Next, we observe in Dec. 2020 a decrease at about 50\% of the intensity contrast with a decrease of the median flux of 7\% regarding Feb. 2020. The environment seems to be smoother compared to Feb. 2020 with a median value getting closer from the Dec. 2018 values. 

Looking at the optical depth map of Dec. 2018 and Dec. 2020, that have a denser (u,v)-plane coverage than Feb. 2020, we can identify opacity variations up to a factor 5 and 3, far more than the estimated 6\% error (section~\ref{sec:varPseudoCont}), respectively. The SiO (2--0) opacity depth map shows therefore strong temporal variations within 2 years, indicative of vigorous changes in the star's environment in this time span.

%% Déplacé en annexe A
%\subsection{Overall optical depth map analysis}
%\textbf{Looking at the optical depth map of 2018-12 and 2020-12, that have a denser (u,v)-plane coverage, we can identify opacity variations up to a factor 5 and 3, far more than the estimated 6\% error (section~\ref{sec:varPseudoCont}), respectively. In order to identify possible artifacts induced by the (u,v)-plane, we used the ASPRO2 software to simulate (u,v) observations with a similar (u,v) plane as for our observations, and used the \texttt{IRBIS} software to reconstruct images of uniform disks. These images (Figure~\ref{fig:IRBIS_test}) are made using the same reconstruction parameters than for the pictures in Figure~\ref{fig:IMAGES}. All the structures identified in Figure~\ref{fig:IRBIS_test} which deviates from the uniform disk model are considered as artifacts. Except from the X's shaped structure observed during the 2020-02 epoch, such structures seems not to be present also in Figure~\ref{fig:IMAGES}, hence we can assume that the variations observed in the reconstructed opacity map at the 2018-12 and 2020-12 epochs are real.}

\subsection{Dust Clump and Cold Spot model}

In addition we checked the robustness of the cold spot and dust clump model used in \citet{Montarges2021} to explain the GDE with respect to our Feb. 2020 VLTI/MATISSE data. Both models are derived from radiative transfer code and own therefore wavelength-dependent structures. Here, we use the exact same models with the same parameters as in \citet[][the reader can refer to that paper to get the models parameters values]{Montarges2021}, just in a different wavelength domain. Thus, the dust cloud and the cold spot presented in this paper as seen by MATISSE in L-band (4000\,nm), may not exhibit the same dynamic range as in \citet{Montarges2021} in V-band ($\approx$650\,nm).\\

\begin{figure}
    \centering
    \includegraphics[width=0.45\textwidth]{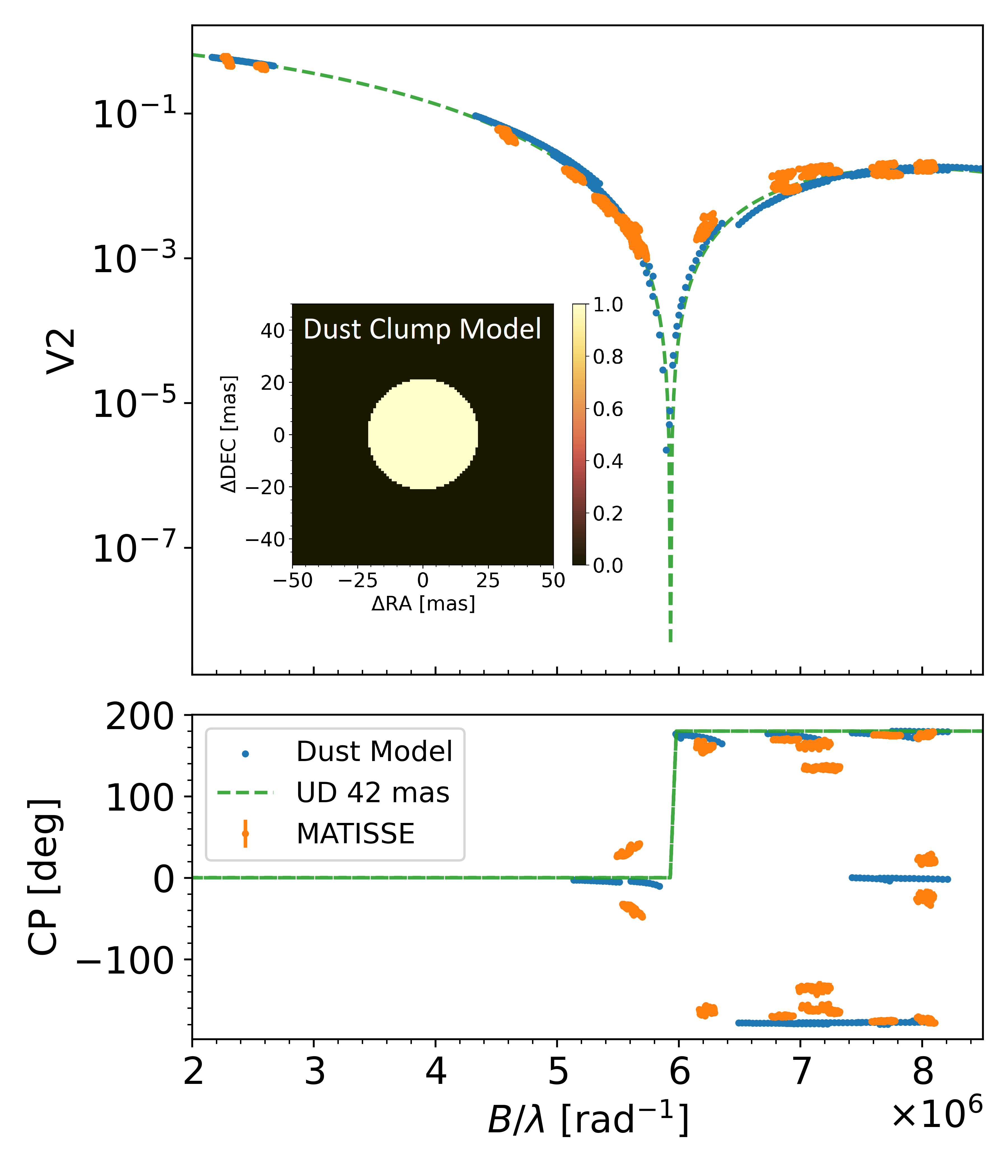}
    \caption{Upper panel: Visibilities squared in the continuum of the MATISSE data \textbf{from Feb. 2020} (orange), dust patch model (blue) with its image in the left corner, and an uniform disk of angular diameter 42\,mas (green). Lower panel: representation of their respective closure phase \textbf{with respect to the spatial frequency of the longest baseline in the baselines triplet.}}\label{fig:models_dust}
\end{figure}

Indeed, in the top panel of Figure~\ref{fig:models_dust} showing the dust clump model, we can see that the dynamic range of the model image (inset) used to compute the L-band visibilities (blue dots) is insufficient to distinguish the models from an uniform disk model (green dashed line). The observed data (orange dots) exhibit small deviation from both the Uniform disk model and the dust clump model. Concerning the closure phase, the dust clump model exhibits a slight deviation from zero or 180$^\circ$ closure phase (i.e. deviation from the centro-symmetric geometry), but not sufficient enough to reproduce the MATISSE data. The same conclusions can be drawn looking at Figure~\ref{fig:models_spot} for the cold spot model. The visibilities are also consistent with the MATISSE data but the closure phase does not seems to show any deviation from zero or 180$^\circ$ closure phase. Thus, both cold spot and dust clump models in L-band are consistent with the MATISSE visibilities but are not able to reproduce the closure phase.

\section{Conclusion}

Our observations suggest an optically thin molecular medium in the SiO (2--0) band absorption at the three epochs. They suggest also the presence of an infrared excess in the pseudo-continuum during the GDE, which has been interpreted as new hot dust formed. We also detect the presence of brighter structures in L-band appearing during the GDE both in the continuum and in the SiO bands. The MATISSE visibilities observations of Fev. 2020 are both compatible with the presence of a cold spot on the surface of the star and that of a cloud of dust formed in the line of sight, but these two models struggle to explain the observed closure phases. The Dec. 2020 observations suggests that Betelgeuse seems to be returning to a gas and surface environment similar to the one observed in Dec. 2018 but with smoother structures maybe due to unusual amount of dust recently formed during the GDE in the line of sight. Opacity maps also show strong spatial variations in Dec. 2018 and Dec. 2020 epochs by respectively a factor of 5 and 3, evidencing a clear inhomogeneous formation of SiO in the stellar atmosphere, as well as vigorous changes in a timespan of 2 years. In addition to the opacity maps, the chromatic estimations of the equivalent uniform disks visibility-fitted angular diameters give also hints to constrain the geometry of the SiO dust seed formation region.

\section*{Acknowledgments}

The authors acknowledge support from the French National Research Agency (ANR) funded project PEPPER (ANR-20-CE31-0002). The results are based on public data released under Programme IDs 60.A9257(D-E), ID 104.20V6, and ID 106.21L4. This work was supported by the \emph{Action Spécifique Haute Résolution Angulaire} (ASHRA) of CNRS/INSU co-funded by CNES, as well as from the \emph{Programme National de Physique Stellaire} (PNPS) from CNRS. It is also supported by the French government through the UCA-JEDI Investments in the Future project managed by the National research Agency (ANR) with the reference number ANR-15-IDEX-01 and the Hungarian government through NKFIH grant K132406. The authors have made extensive use of the Jean-Marie Mariotti Center (JMMC) tools. 

{\bf Data availability:} The data underlying this article are available in the OIdb digital repository with direct links to the data collections: \href{http://oidb.jmmc.fr/collection.html?id=5866421b-f263-46c0-a642-1bf263a6c0bd}{Dec. 2018}, \href{http://oidb.jmmc.fr/collection.html?id=6c5e6803-f7af-4f0e-9e68-03f7afcf0e18}{Feb. 2020}, and \href{http://oidb.jmmc.fr/collection.html?id=07f92587-f7ef-4f04-b925-87f7ef3f04c1}{Dec. 2020}.

\bibliographystyle{mnras}
\bibliography{betelgeuse}

%\newpage

\appendix

\section{Image reconstruction of an uniform disk with IRBIS}
\label{sec:simIRBIS}

\begin{figure*}
    \centering
    \includegraphics[width=1\textwidth]{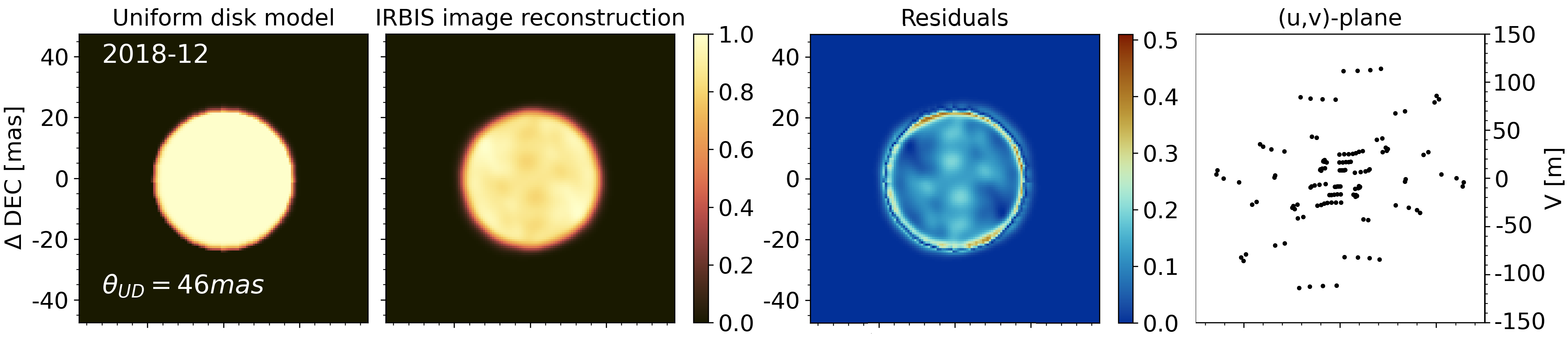}
    \includegraphics[width=1\textwidth]{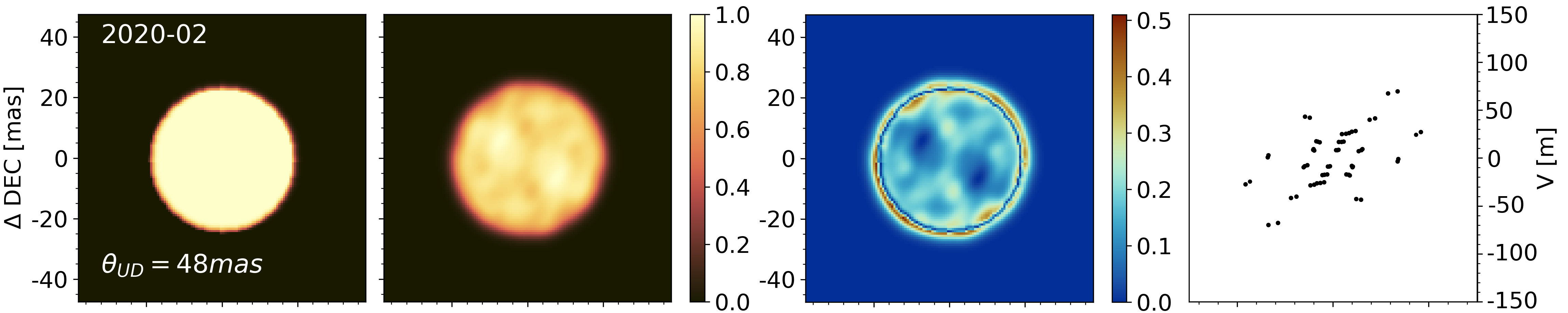}
    \includegraphics[width=1\textwidth]{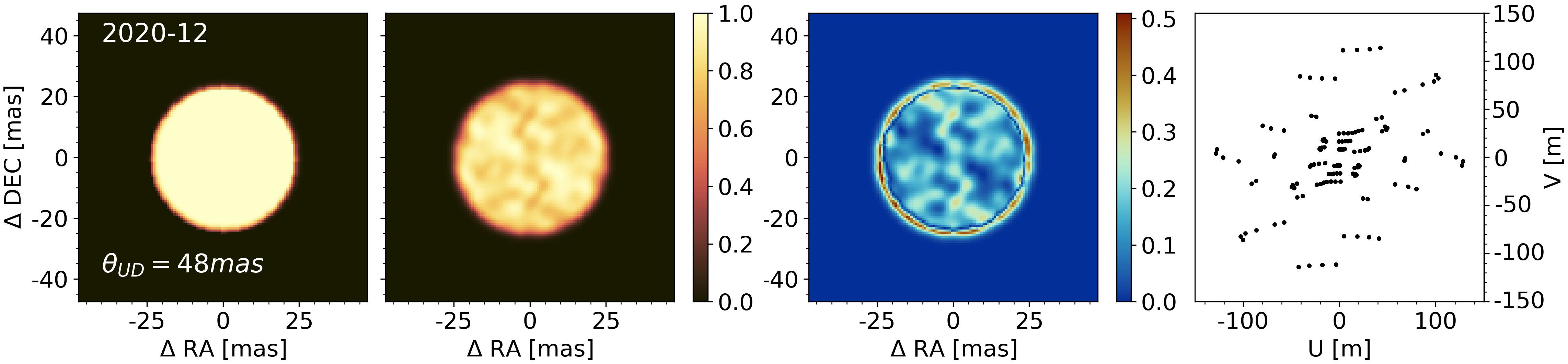}
    
    \caption{Each row correspond to a given epoch precised in the top-left corner of the first panel. From left to right: 1) model of the uniform disk convolved with the same interferometric beam as used in Figure~\ref{fig:IMAGES}, 2) reconstructed image using \texttt{IRBIS} of a simulated interferometric data of an uniform disk with an angular diameter determined using the values fitted on the visibility squared showed in Figure~\ref{fig:FIT} for the SiO first overtone wavelength, 3) residuals between the model and the image reconstructed, 4) simulated (u,v)-plane coverage used for the image reconstruction of the simulated data.} \label{fig:IRBIS_test}
\end{figure*}

In order to identify possible artifacts in our images induced by the (u,v)-plane, we used the ASPRO2 software to simulate (u,v) observations with a similar (u,v) plane as for our observations, and we used the \texttt{IRBIS} software to reconstruct images based on these simulated data. These images (Figure~\ref{fig:IRBIS_test}) are made using the same reconstruction parameters as for the pictures in Figure~\ref{fig:IMAGES}. All the structures identified in Figure~\ref{fig:IRBIS_test} which deviates from the uniform disk model are considered as artifacts. Except from the X-shaped structure observed during the Feb. 2020 epoch, the symmetric structures seen in our simulation  do not seem to be present in Figure~\ref{fig:IMAGES}, hence we can assume that the variations observed in the reconstructed opacity map at the Dec. 2018 and Dec. 2020 epochs are real.

%{\bf In order to identify possible artifacts induced by the (u,v)-plane, we used the \texttt{IRBIS} reconstructed images of uniform disks of Figure~\ref{fig:IRBIS_test} made using the same reconstruction parameters than for the pictures in Figure~\ref{fig:IMAGES}. All the structures identified in Figure~\ref{fig:IRBIS_test} which deviates from the uniform disk model are considered as artifacts. Except from the X's shaped structure observed during the 2020-02 epoch, such structures seems not to be present also in Figure~\ref{fig:IMAGES}, hence we can assume that the variations observed in the reconstructed opacity map at the 2018-12 and 2020-12 epochs are real.}

\section{Optical depth map estimation} \label{sec:depth_map}

From these images we derive the spectral distribution of the optical depth at the SiO absorption bands $\tau_{\rm SiO}(\lambda)$ assuming that: 
\begin{enumerate}[label=\arabic*)]
\item the stellar photosphere is the dominant source of the pseudo-continuum radiation, denoted herafter as $S_{\star}(\lambda)$; 
\item the SiO material is responsible not only for the absorption but also for the emission of the light in the SiO absorption bands, denoted hereafter as $S_{\rm SiO}(\lambda)$. 
%\item the SiO gas cloud also contributes in emission through the $S_{\rm SiO}(\lambda)$ function. \PC{re-emits absorbed light ???} \JD{Ben oui la lumière absorbée et bien ré-émise par le nuage d'une façon ou d'une autre, après tout est une question de savoir si l'émission du nuage est majoritaire par rapport à son absorption, c'est ce que te montre la formule du transfert radiatif, dans notre cas de figure puisque nous avons des raies d'absorption, on voit bien que la partie émissive n'est pas majoritaire et c'est ce que l'on dit plus loin dans les hypothèses. Mais si on veut rester général il faut parler des deux contributions dans un premier temps.} \PC{je trouve que c'est pas très bien dit, je vais essayer de reformuler n'enlève pas ce que tu as écrit stp, merci !}.
\end{enumerate}
% We have developed a technique to extract the optical depth map from the image reconstructions. Let us describe it.
% In this work, we can assume that the SiO shell is homogeneous and sits right in front of the star. Our setup is therefore composed of two object: the star's photosphere and the SiO shell. In the continuum, we assume that only the stellar photosphere contributes to the observed intensity. In the line, 3 contributions are present: the  stellar photosphere (stellar source $S_{\star}(\lambda)$), the SiO source ($S_{\rm SiO}(\lambda)$), and the SiO  opacity $\tau_{\rm SiO}(\lambda)$.
Under these assumptions, the radiative transfer equation becomes:

\begin{equation}
I(\lambda) = S_{\star}(\lambda)\,e^{-\tau_{\rm SiO}(\lambda)}+S_{\rm SiO}(\lambda) \left[ 1-e^{-\tau_{\rm SiO}(\lambda)} \right] ,
\end{equation}

where $I(\lambda)$ is the spectral distribution of surface brightness. Since the absorption feature in the spectrum is weak (c.a. 20\% of the pseudo-continuum), we assume that the SiO layer is optically thin, i.e. $\tau_{\rm SiO}(\lambda_{\rm band}) < 1$.

%\PS{To estimate the SiO optical depth at $\lambda_{\rm line}$ using the image reconstructions, we assume several hypotheses to simplify the extraction of opacity maps:}
In order to get an estimate of the optical depth, we make the following additional simplifying assumptions:
\begin{enumerate}[label=\arabic*)]
 \setcounter{enumi}{2}

    \item the SiO contribution to the continuum wavelength range is negligible i.e. $\tau_{\rm SiO}(\lambda_c) = 0$, hence:
\begin{equation}
I(\lambda_c) = S_{\star}(\lambda_c); \label{eq:Icont}
\end{equation}
    \item in front of the stellar disk, the SiO emission is negligible compared to the stellar emission in the absorption band, so that: 
\begin{align}
&S_{\rm SiO}(\lambda_{\rm band}) \ll S_{\star}(\lambda_{\rm band}),~\rm{and}\\
&S_{\rm SiO}(\lambda_{\rm band}) \left[1-e^{-\tau_{\rm SiO}(\lambda_{\rm band})} \right] \approx 0 ,
\end{align}
hence:
\begin{equation}
I(\lambda_{\rm band}) = S_{\star}(\lambda_{\rm band})\,e^{-\tau_{\rm SiO}(\lambda_{\rm band})}. \label{eq:Iline}
\end{equation}

\begin{comment}
Finally, we assume that the stellar flux is blackbody-like ($S_{\star}(\lambda) = B_{\star}(\lambda, T_\star)$) and that the change in wavelength between line and continuum does not significantly change the continuum level i.e.
\begin{equation}
B_{\star}(\lambda_c, T_\star) = B_{\star}(\lambda_{\rm band}, T_\star) = B_{\star}.
\end{equation}
In practice, the flux for a blackbody with $T_\star=3500\,K$ changes by $\leq$ 2\% between 3.942$\mu$m and 4.012$\mu$m, much less than the variation seen with position in Fig.~\ref{fig:IMAGES}.
\end{comment}

    \item the stellar source function is the same in the absorption band as in the continuum as long as the continuum level does not significantly change between both spectral channels, so that:

\begin{equation}
S_{\star}(\lambda_c) = S_{\star}(\lambda_{\rm band}).    
\end{equation}
\end{enumerate}

Substituting these assumptions into equations \ref{eq:Icont} and \ref{eq:Iline} leads to a rough estimate of the optical depth at the wavelength range of the SiO (2--0) absorption band :
\begin{equation}
\tau_{\rm SiO}(\lambda_{\rm band}) = \log I(\lambda_c) - \log I(\lambda_{\rm band}).
\end{equation}

\section{Cold Spot Model}

\begin{figure}
    \centering
    \includegraphics[width=0.45\textwidth]{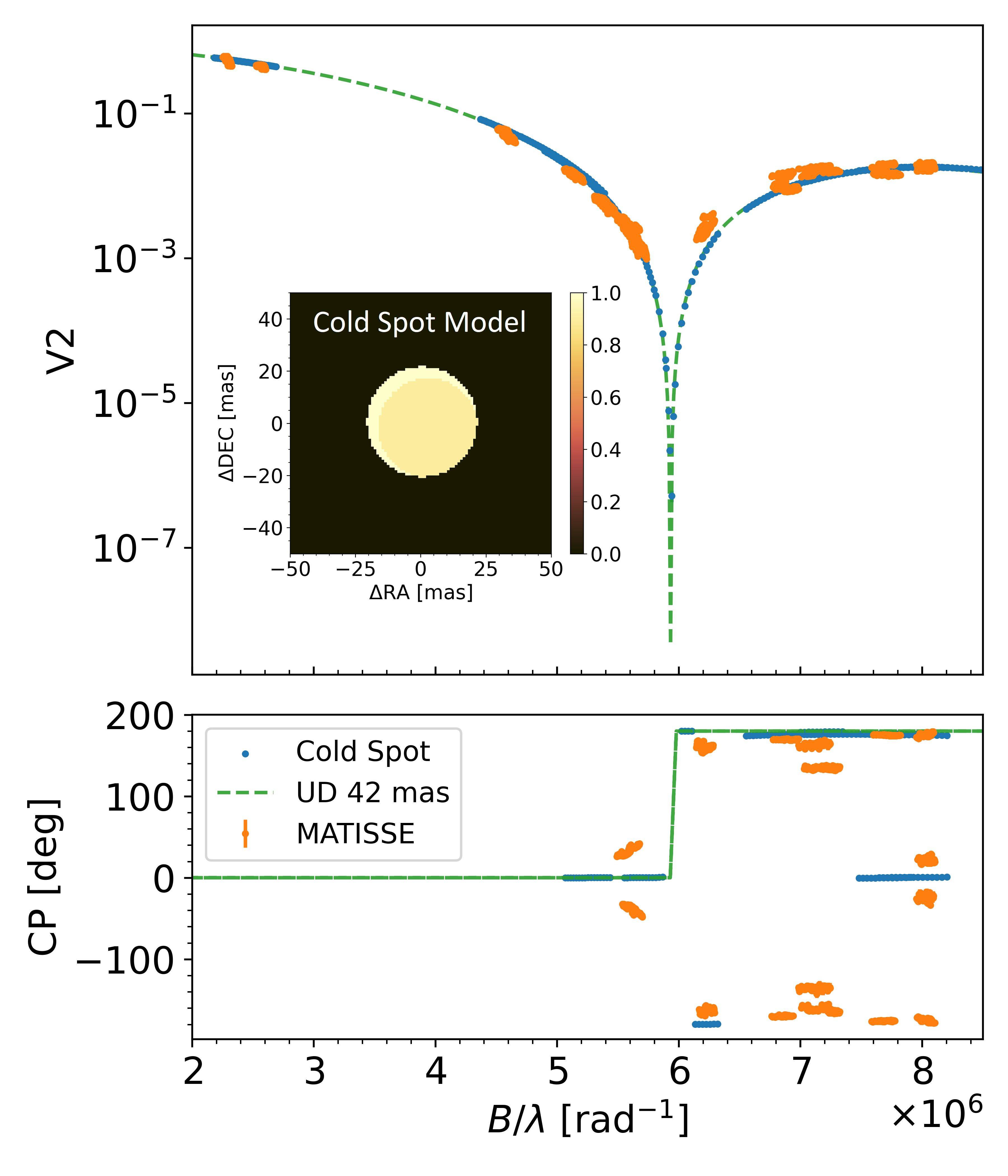}
    \caption{Upper panel: Visibilities Squared in the continuum of the MATISSE data \textbf{of February 2020} (orange), cold spot model (blue) with its image in the left corner, and an uniform disk of angular diameter 42\,mas (green). Lower panel: representation of their respective closure phase.   }\label{fig:models_spot}
\end{figure}

%\end{adjustwidth}

\end{document}